\documentclass[useAMS,usenatbib,usegraphicx]{mn2e}

\title[NGC~1407 and NGC~1400: kinematics and photometry]
{The early-type galaxies NGC~1407 and NGC~1400 $-$~I:
spatially resolved radial kinematics and surface photometry}

\author[M. Spolaor et al.]{Max Spolaor$^1$\thanks{E-mail: mspolaor@astro.swin.edu.au}, 
Duncan A. Forbes$^1$, George K. T. Hau$^{1,2}$, Robert N. Proctor$^1$,\newauthor and Sarah Brough$^1$\\
$^1$Centre for Astrophysics \& Supercomputing, Swinburne University, Hawthorn, VIC 3122, Australia\\ 
$^2$Department of Physics, University of Durham, South Road, Durham, DH1 3LE, UK\\}

\begin{document}
\date{Accepted... Received...; in original form 2007}
\pagerange{\pageref{firstpage}--\pageref{lastpage}} \pubyear{2007}

\maketitle
\label{firstpage}

\begin{abstract}
This is the first paper of a series focused on investigating the star formation and evolutionary history of the two early-type galaxies NGC~1407 and NGC~1400. They are the two brightest galaxies of the NGC~1407 (or Eridanus-A) group, one of the 60 groups studied as part of the Group Evolution Multi-wavelength Study (GEMS). 

Here we present new high signal-to-noise long-slit spectroscopic data obtained at the ESO 3.6m telescope and high-resolution multi-band imaging data from the HST/ACS and wide-field imaging from Subaru Suprime-Cam. We spatially resolved integrated spectra out to $\sim$~0.6 (NGC~1407) and $\sim$~1.3 (NGC~1400) effective radii. The radial profiles of the kinematic parameters $v_{rot}$, $\sigma$, $h_{3}$ and $h_{4}$ are measured. The surface brightness profiles are fitted to different galaxy light models and the colour distributions analysed. The multi-band images are modelled to derive isophotal shape parameters and residual galaxy images. The parameters from the surface brightness profile fitting are used to estimate the mass of the possible central supermassive black hole in NGC~1407. The galaxies are found to be rotationally supported and to have a flat core in the surface brightness profiles. Elliptical isophotes are observed at all radii and no fine structures are detected in the residual galaxy images. From our results we can also discard a possible interaction between NGC~1400, NGC~1407 and the group intergalactic medium. We estimate a mass of $\sim$1.03~$\times$~10$^{9}$~$M_{\sun}$ for the supermassive black hole in NGC~1407 galaxy.
\end{abstract}


\section{INTRODUCTION}
Early-type galaxies are estimated to contribute to at least half, and perhaps as much as three quarters, of the stellar mass in the local Universe \citep{bell03}. Understanding their formation and evolution is therefore a fundamental objective of astrophysical research. 

\begin{table*}
\begin{center}
\begin{tabular}{ccccccccccc}
\hline 
\hline 
Name & R.A. & DEC. & Type &PA & v & $\sigma_{0}$ &$r_{e}$ & M$_{B}$ & L$_{X}$\\ 
       &(J2000) & (J2000) & &(degrees) &(km s$^{-1}$)& (km s$^{-1}$)& (arcsec) &(mag) & (erg~s$^{-1}$) \\ 
\hline 
NGC~1407 & 03:40:11.9 & $-$18:34:49 & E0 &35 &1779 $\pm$ 9 & 305.0 $\pm$ 10.0 &72$\arcsec$ & $-$21.22 & 1.00~$\times$~10$^{41}$ \\ 
NGC~1400 & 03:39:30.8 & $-$18:41:17 & E$-$S0 & 40 & 558 $\pm$ 14 & 279.0 $\pm$ 2.0 &27$\arcsec$ & $-$19.95 & 1.32~$\times$~10$^{40}$ \\ 
\hline 
\hline
\end{tabular}
\end{center}
\caption{Galaxies properties. Type, P.A.: morphological type and position angle of the major axis from HyperLEDA; v, $\sigma_{0}$, $r_{e}$: velocity, central velocity dispersion and effective radius from HyperLEDA; M$_{B}$: absolute B-band magnitude obtained from HyperLEDA; L$_{X}$: X-ray luminosity from \citep{osullivan01}.} 
\label{ga_prop}
\end{table*}

The debate on whether early-type galaxies are formed by monolithic collapse at high redshifts or via hierarchical merging is still not settled. Photometrically, the tightness of the colour-magnitude
diagram suggests that the bulk of the stars were formed at high redshifts, z~$>$~1 (e.g \citealt{bower92}; \citealt{kodama99}). However, signs of recent mergers and interactions are often present when images of early-type galaxies are examined (e.g. \citealt{ft92}). Shells and ripples are believed to be formed by mergers (e.g. \citealt{quinn84}) or by interactions (e.g. \citealt{thomson90}; \citealt{thomson91}). Deviations from perfect elliptical isophotes are correlated with the galaxy shape and amount of rotation support (e.g. \citealt{khoc05}). Furthermore, boxyness is often associated with X-ray and radio activities which may indicate an evolutionary history more affected by mergers (e.g. \citealt{bender89}). Hence the detailed photometric study of early-type galaxies may yield important clues on their past merger activities.

Clues to the formation of early-type galaxies may also come from their kinematics. Studies on the global kinematics have established that elliptical galaxies as a class are supported by velocity anisotropy
(e.g. \citealt{binney76}; \citealt{binney78}). Detailed kinematic studies often reveal kinematic distinct cores (e.g. \citealt{emsellem04}) and non-relaxed structure (e.g. \citealt{balce99}) which may be related to the way the galaxy is formed and to the merger history. It is possible to measure the shape of absorption lines, hence the Line-Of-Sight Velocity Distribution (LOSVD), which will tell us about the velocity anisotropy and hence constrain the orbital families (\citealt{bn90}; \citealt{rix92}; \citealt{van93}). The LOSVD is often parametrised by the mean velocity $v_{rot}$ and velocity dispersion $\sigma$, plus higher order moments ($h_{3}$, $h_{4}$ ....) of a Gauss-Hermite series. The $h_{3}$ and $h_{4}$ offer extra information on the asymmetric and symmetric deviation, respectively, away from a perfect Gaussian.

Observationally, we now have the ability to obtain spatially resolved high signal-to-noise integrated spectra out to large galactic radii and hence larger mass fractions. The high-resolution multi-band imaging data provide further insights on the radial distribution of the galaxy light and allow meaningful comparisons with spectroscopically derived results. Therefore, we are able to consider physical mechanisms acting locally and to investigate how their properties vary with the galactocentric radius.

In this series of papers we focus on the internal properties of the two early-type galaxies NGC~1407 and NGC~1400 in an effort to understand their star formation and evolutionary history. In this first paper (hereafter Paper~I), we present spatially resolved radial profiles of the kinematic parameters $v_{rot}$, $\sigma$, $h_{3}$ and $h_{4}$ out to $\sim$~0.6 (NGC~1407) and $\sim$~1.3 times (NGC~1400) the galaxies' effective radii ($r_{e}$; the radius within which half the galaxy light is contained). We also present spatially resolved multi-band high-resolution wide-field imaging data out to $\sim$~1.4$r_{e}$ for both galaxies. The spatial distributions of the galaxy light profiles are analysed and the isophotal ellipticity radial profiles and deviation parameters recovered. The residual images of the two galaxies are inspected for fine structure (e.g. dust, ripples, tidal tails, boxy or disky structure, etc.). In the second paper (\citealt{spola07b}, hereafter Paper~II) we will present spatially resolved stellar population parameters from the same high-S/N long-slit spectroscopic data used here. The aim of Paper~II will be to interpret and combine the results from the stellar populations analysis with the kinematic and photometry results.

This paper is organised as follows. In Section~2 we describe the two sample galaxies. In Sections~3 we describe the spectroscopic and photometric observations, together with an explanation of the relevant data reduction procedures. Section~4 presents the spatially resolved radial kinematic profiles. Section~5 presents the spatially resolved surface photometry results. In Section~6 we discuss a possible interaction between NGC~1400 and NGC~1407 and the group intergalactic medium (IGM). In Section~7 we summarise our results.

\section{THE DATA SAMPLE}

\begin{table}
\begin{center}
\begin{tabular}{cccc}
\hline 
\hline 
Method                  & NGC 1407        & NGC 1400 &   $\Delta D$  \\
and Reference           & Distance        &Distance  &               \\
                        & (Mpc)           & (Mpc)    & (Mpc)         \\
\hline
F$-$J (1) & 18.9 $\pm$ 1.6  & 31.5 $\pm$ 2.9 & $-$12.6 $\pm$ 3.3 \\
Virgo ICV (1) & 22.6 $\pm$ 0.2  &  5.7 $\pm$ 0.2 & $+$16.9 $\pm$ 0.3 \\
$D_{n} - \sigma$ (2)  & 26.4 $\pm$ 1.2  & 33.3 $\pm$ 1.2 & $-$6.9 $\pm$ 1.7  \\
L$_{X}$ $-$ L$_{B}$ (3)  & 25.0 $\pm$ 1.4  & 22.2 $\pm$ 2.8 & $+$2.8 $\pm$ 3.1 \\
SBF             (4)  & 16.4 $\pm$ 0.9  & 16.3 $\pm$ 1.0 & $+$0.1 $\pm$ 1.3  \\
GCLF            (5)  & 17.6 $\pm$ 3.1  & 25.4 $\pm$ 7.0 & $-$7.8 $\pm$ 7.6  \\
SBF             (6)  & 28.8 $\pm$ 3.7  & 26.4 $\pm$ 4.4 & $+$2.4 $\pm$ 5.7  \\
SBF             (7)  & 26.8 $\pm$ 1.6  & 24.5 $\pm$ 4.1 & $+$2.3 $\pm$ 2.8  \\
SBF             (8)  & 25.1 $\pm$ 1.2  &  $-$           & $-$   \\
GCLF            (9) & 20.9 $\pm$ 1.0  &  $-$           & $-$   \\
\hline
\textbf{Average} & 22.8 $\pm$ 1.9  & 23.2 $\pm$ 3.6 & $-$0.4 $\pm$ 4.1 \\
\hline
\textbf{Adopted}& 21.0            & 21.0           & 0.0 \\
\hline
\hline
\end{tabular}
\end{center}
\caption{Distance measurements of the galaxies NGC 1407 and NGC 1400. SBF: surface brightness fluctuations. GCLF: Globular cluster luminosity function. F$-$J: Faber$-$Jackson relation (i.e. $m - M = B_{T} + 6.2  log \sigma + 5.9$). Virgo ICV: galaxy velocity corrected for infall of the Local Group towards Virgo. (1): HyperLEDA (2): \citet{faber89}. (3): \citet{donnelly90}. (4): \citet{tonry91}. (5): \citet{perrett97}. (6): \citet{tonry01}. (7): \citet{jensen03}. (8): \citet{cantiello05}. (9): \citet{forbes06b}.}
\label{distance}
\end{table}
The NGC~1407, or Eridanus-A, group is one of the 60
groups studied as part of the Group Evolution Multi-wavelength Study (GEMS; \citealt{osmond04}; \citealt{forbes06}). The study was designed to probe the evolution of groups and their member galaxies. The NGC~1407 group is dominated by the NGC~1407 galaxy. It contains the intermediate luminosity (M$_{B}~\leq$~$-$19.65~mag) early-type galaxies NGC~1440, NGC~1452, NGC~1393, NGC~1383, IC~346 and NGC~1359 and a large number of dwarfs, recently estimated by \cite{trentham06} to number 250. \cite{brough07} used velocities for 25 member galaxies to study the group dynamics. They found a mean recession velocity $v$~=~1652~km~s$^{-1}$, a velocity dispersion $\sigma_{v}$~=~384~$\pm$~63~km~s~$^{-1}$ and estimated the mass of the group to be 7.9~$\times$~10$^{13}$M$_{\odot}$ with a M/L$_{K}$~=~230M$_{\odot}$/L$_{\odot}$. Based on 35 measurements, \cite{trentham06} estimated, in good agreement, $v$~=~1630~km~s$^{-1}$, $\sigma_{v}$~=~387~$\pm$~65~km~s~$^{-1}$ and a mass of 7.3~$\times$~10$^{13}$M$_{\odot}$ with M/L$_{R}$~=~340M$_{\odot}$/L$_{\odot}$. 

The early-type galaxies NGC~1407 and NGC~1400 (see Table~\ref{ga_prop} for details) lie at small projected angular distance on the sky ($\sim$~12~arcmin) but with a remarkable difference in velocity, $\Delta v \sim$~1100~km~s$^{-1}$. NGC~1407 is the brightest group galaxy, whereas NGC~1400 is the second brightest galaxy with a peculiar velocity of 558~km~s$^{-1}$. Various independent studies have attempted to measure the galaxies' distances using different techniques. A summary is given in Table~\ref{distance}. The results show that the galaxies lie at almost the same distance, despite the difference in velocity. In this work we assume a distance modulus of 31.6 for both galaxies, which for $H_{0}$~=~72~km~s$^{-1}$~Mpc$^{-1}$ ($\Omega_{m}$ = 0.3, $\Omega_{\Lambda}$ = 0.7) implies a distance of 21 Mpc where 1 arcmin corresponds to 6.12 kpc (\citealt{bender92}; \citealt{forbes06b}).


\section{OBSERVATIONS AND DATA REDUCTION}
\subsection{ESO/EFOSC2 spectroscopic data}
Spectral observations were performed with the ESO Faint Object Spectrograph
and Camera (EFOSC2) mounted on the ESO 3.6m telescope at La Silla
Observatory, Chile.  Data for NGC~1407 and NGC~1400 were collected
during an observing run on 2004 Dec.~11-12 during which other early-type
galaxies were also observed (see \citealt{hau06}). Table~\ref{obs_conf} 
summarises the instrumental configuration adopted during the observing run. Lick/IDS and spectrophotometric standard stars were taken at the
parallactic angle, the latter with a 5$\arcsec$ wide slit. The Lick/IDS stars are also used
as velocity standards.

Data reductions were carried out with IRAF\footnote{IRAF is
distributed by the National Optical Astronomy Observatories, which are
operated by the Association of Universities for Research in Astronomy,
Inc., under cooperative agreement with the National Science
Foundation.} using the procedures adopted in
\cite{hau06}. The data were bias-subtracted and flat-corrected. The
spectra were re-binned to logarithmic wavelength after correction
for instrumental response and spatial distortion. External regions
of the 2D spectrum were used to measure the sky background and to
subtract it from the final data. The three exposures were co-added
before the extraction of the 1D spectra. 

We spatially resolved 68 and 82 apertures along the observing axes for
NGC~1407 and NGC~1400, respectively. The spatial width (i.e. the number of CCD rows binned) for each
extracted aperture increases with radius to achieve a signal-to-noise
ratio of $\sim$~30~{\AA}$^{-1}$ at 5000~{\AA}. We reached a radial extent of $\sim$~0.56$r_{e}$ ($\sim$~4.11 kpc) for NGC~1407 and $\sim$~1.30$r_{e}$ ($\sim$~3.58 kpc) for NGC~1400.

\begin{table}
\begin{center}
\begin{tabular}{lc}
\hline 
\hline 
Telescope                   & ESO-3.6m         \\
Spectrograph                & EFOSC2           \\
Chip size (pixels)          & 2048 x 2048      \\
Grism resolution  FWHM      & 7.8 {\AA}         \\ 
Pixel scale                 & 0.314$\arcsec$  \\
Binning                     & 2 x 2            \\
Slit  width                 & 1.2$\arcsec$          \\ 
Wavelength coverage         & 4320 - 6360 {\AA} \\ 
Seeing FWHM                 & $\sim$ 1$\arcsec$    \\
Exposure time               & 3 x 1200 s        \\
NGC~1407 Slit P.A.          & 44$^{\circ}$     \\
NGC~1400 Slit P.A.          & 42$^{\circ}$     \\
\hline 
Telescope                   & HST \\
Camera                      & ACS \\
Pixel Format                & two 2048 $\times$ 4096 CCDs\\ 
Field of View               & 3.5$\arcmin$ $\times$ 3.5$\arcmin$\\
Pixel Scale                 & 0.049$\arcsec$                 \\
Gap size                    & $\sim$ 50 pixels = 2.5$\arcmin$\\
F435W exposure time          & 1500s     \\
F814W exposure time         & 680s      \\
\hline
Telescope                   & Subaru     \\
Camera                      & Suprime-Cam \\
Pixel Format                & ten 2048 $\times$ 4096 CCDs\\ 
Field of View               & 34$\arcmin$ $\times$ 27$\arcmin$\\
Pixel Scale                 & 0.20$\arcsec$               \\
Seeing FWHM                 & $\sim$ 0.5$\arcsec$       \\
g' exposure time          & 240s     \\
r' exposure time         &  120s      \\
i' exposure time         &  60s      \\
\hline
\hline
\end{tabular}
\end{center}
\caption{Observing parameters. Spectral (top table) and imaging (middle and bottom table) observing characteristics.}
\label{obs_conf}
\end{table}

\subsection{HST/ACS imaging data}
The NGC~1407 imaging observations were performed with the Advanced
Camera for Surveys (ACS) mounted on the Hubble Space
Telescope (HST). The images were obtained in the F435W (B) filter and F814W (I)
filter, with an exposure time of 1500s and 680s respectively. To
avoid the gap between the detectors the galaxy centre was shifted
about 1 arcmin from the centre of the ACS field-of-view. The data 
were downloaded from the HST archive (proposal ID = 9427; PI
= Harris). Table~\ref{obs_conf} reports the instrumental
characteristics.

The images were processed by the ACS calibration pipeline (CALACS);
bias-subtraction, dark-subtraction, flat-correction, cosmic ray
rejection and geometric distortion correction were performed. 

The high instrumental resolution of the HST/ACS allowed us to sample NGC~1407 over radii $\sim$~0.1~$-$~100~arcsec ($\sim$~1.39$r_{e}$), corresponding to $\sim$~0.01~$-$~10.20~kpc.

\subsection{Suprime-Cam imaging data}
\label{subaru}
The NGC~1400 imaging was obtained with the Subaru Prime Focus Camera (Suprime-Cam; \citealt{miya02}) located at the prime focus of the Subaru Telescope, which covers a field of view of 34$\arcmin$~$\times$~27$\arcmin$. The data are from \cite{spit07}. The SDSS photometric system filters g', r', i' were used, with exposure times of 240s, 120s and 60s respectively. Table~\ref{obs_conf} reports the instrumental characteristics.

The images were processed by the automatic Data Reduction Software for Subaru Suprime-Cam (SDFRED); bias-subtraction, dark-subtraction, flat-correction, overscan correction, cosmic ray
rejection, geometric distortion and atmospheric dispersion corrections were performed along with masking of bad data, mosaic alignment and final image combination.

The images of the galaxy central regions showed bleeding caused by saturation within the CCD. Consequently, the sampled radii range from $\sim$~5 to 100 arcsec ($\sim$~3.70$r_{e}$), corresponding to $\sim$~0.51~$-$~10.20~kpc. The large field of view covered by the observations also included the NGC~1407 galaxy. For future reference (see Paper~II) we extracted radial colour profiles of NGC~1407 in the range $\sim$~5 to 100 arcsec ($\sim$~1.39$r_{e}$; $\sim$~0.51~$-$~10.20~kpc). 


\section{SPATIALLY RESOLVED RADIAL KINEMATICS}
\label{kinematic}
\subsection{Kinematic analysis}
The description of the kinematic analysis methodology is presented in
\cite{hau06}. Here we give a brief review. 

We used the \cite{van94} code which is based on the method for the
identification of non-Gaussian line profiles in elliptical galaxies
(\citealt{van93}). It parametrises line profiles as a sum of orthogonal
functions in a Gauss-Hermite series; relaxing the hypothesis of a
single Gaussian profile. Best-fitting line-of-sight mean velocity, $v_{rot}$, and
velocity dispersion, $\sigma$, are calculated. Asymmetric (skewness) and
symmetric (kurtosis) deviations in the LOSVD from a simple Gaussian distribution are described
by the parameters $h_3$ and $h_4$, respectively. In a galaxy with large-scale kinematic coherence the $h_{3}$ radial profile is expected to be point-symmetric with respect to the galaxy centre, whereas the $h_{4}$ profile is expected to show an overall symmetry.

A reiterated fitting procedure was applied to improve the fits. Bad pixels, hidden emission lines and bright pixels due to
sky emission were excluded by the process. A list of reference stars were
observed to minimise the template mismatch for the recovered
kinematics: after making a comparison of $\chi^{2}$ values obtained by
different fitting the best template was chosen. 

\subsection{NGC~1407 results}
\begin{figure*}
\includegraphics[scale=.5]{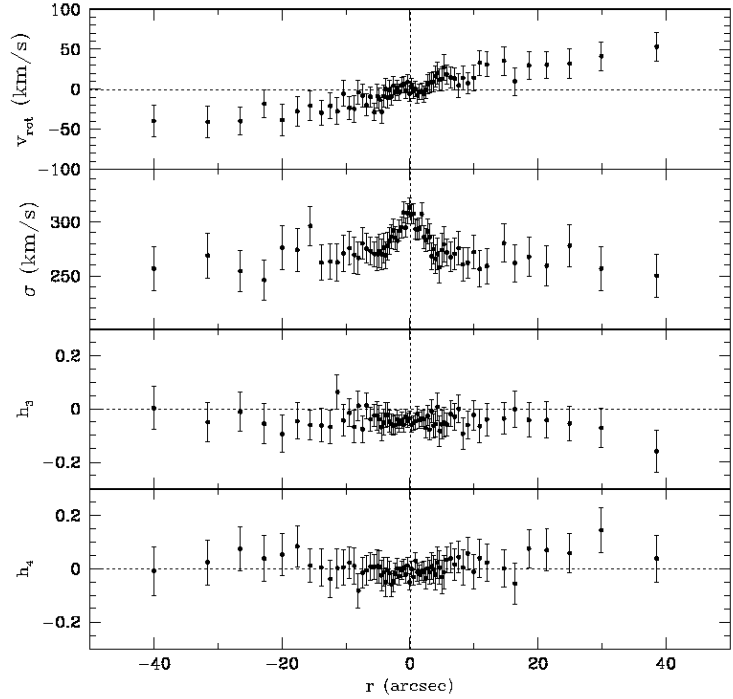}
\caption[]{Large-scale major-axis kinematics of NGC~1407. In each panel, from top to bottom are shown rotation velocity, velocity dispersion, third and fourth Hermite terms. The last two parameters quantify deviations from a simple Gaussian velocity distribution (\citealt{van94}). The rotation velocity profile shows a possible kinematically decoupled core in the central 5 arcsec (0.51 kpc).}
\label{kin1}
\end{figure*}

Several previous works (e.g. \citealt{davies87}; 
\citealt{faber89}; \citealt{beuing02}) have studied the kinematic properties of the NGC 1407 central regions. \cite{longo94} and \cite{franx89} obtained spatially resolved spectra out to radii of 10 arcsec ($\sim$~0.14$r_{e}$) and 30 arcsec ($\sim$~0.42$r_{e}$), respectively. \cite{franx89} analysed the largest radius but with few data points. In contrast, the radial range in \cite{longo94} is well-sampled, but the maximum radius is only a small fraction of the effective radius. The data used in our analysis increase the spatial coverage, up to $\sim$~0.56$r_{e}$, and allow us to spatially resolve 68 regions along the major axis. The radial profiles are presented in Figure~\ref{kin1} and the kinematic measurements are reported in Table~\ref{kinematics} (all the data are available electronically).

\begin{figure*}
\includegraphics[scale=.5]{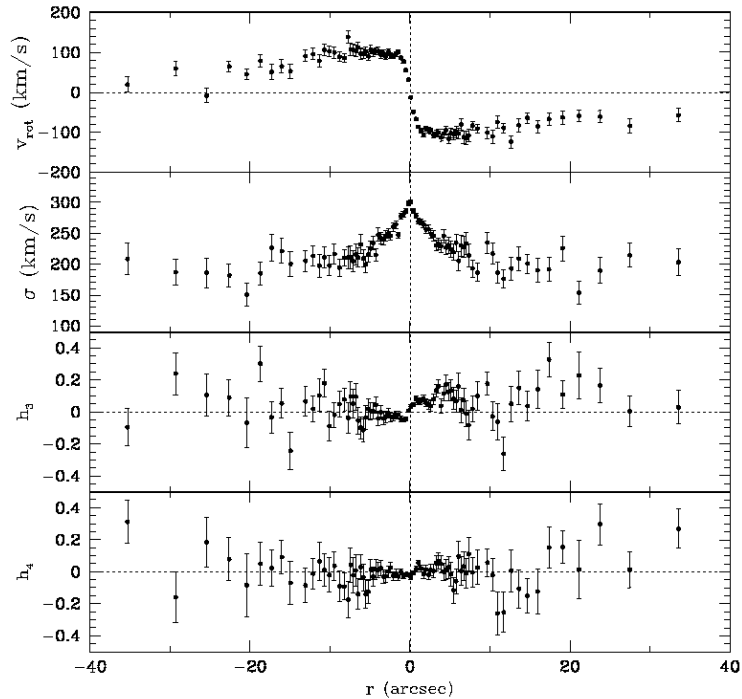}
\caption[]{Large-scale major-axis kinematics of NGC~1400. Description as in Fig.~\ref{kin1}.}
\label{kin2}
\end{figure*}

The velocity profile (Fig.~\ref{kin1}) shows little rotation in the outer region, $v_{max} \leq$~50~km~s$^{-1}$, and the possible signature of a kinematically decoupled core (KDC) in the central 5 arcsec (0.51 kpc). This peculiar feature implies that at small radii the stars rotate in the opposite direction to the main body of the galaxy. The velocity dispersion profile is sharply peaked at the galaxy centre reaching a maximum value of 310~km~s$^{-1}$. The $h_{3}$ profile appears to be affected by a systematic downward-offset. Performing various tests during the reduction procedure we found that the $h_{3}$ offset could be caused by a template mismatch. A small slit misalignment with respect to the galaxy nucleus might also explain the offset. This is because, we might be sampling stars which are streaming towards or away from the line-of-sight, so the line profile will be skewed especially in the centre. In summary, within errors the $h_{3}$ and $h_{4}$ profiles are flat and do not allow a confirmation of the KDC.

The anisotropy parameter $(v/ \sigma)^{*}$ (\citealt{binney78}; \citealt{bn90}) is defined to be the ratio of observed $v_{max}/\sigma_{0}$ to that expected for an oblate isotropic rotating galaxy of same ellipticity. It is independent of ellipticity and projection effects, so it may be used to separate galaxies that are rotationally supported ($(v/\sigma)^{*}\leq$~0.7; \citealt{bender92}) from those that are anisotropic (supported by the pressure of anisotropic stellar orbits).
We found NGC~1407 to be marginally rotationally supported with a ratio of 0.74~$\pm$~0.13.

\begin{table}
\begin{center}
\begin{tabular}{ccc}
\hline   
\hline
        & NGC 1407 & NGC 1400 \\
\hline
$v_{system}$             & 1794.0 $\pm$ 9.8 km s$^{-1}$ & 560.0 $\pm$ 10.4 km s$^{-1}$\\
$v_{max}$                & 53.4 $\pm$ 18.5  km s$^{-1}$& 138.5 $\pm$ 16.5 km s$^{-1}$\\
$\sigma_{0}$             & 313.4 $\pm$ 9.2  km s$^{-1}$& 300.6 $\pm$ 3.7 km s$^{-1}$ \\
$(v/\sigma_{0})^{*}$                & 0.74 $\pm$ 0.13  & 1.18 $\pm$ 0.12  \\
\hline 
\hline
\end{tabular}
\end{center}
\caption[]{Kinematic measurements. The systemic velocity $v_{system}$ is taken as the mean of all the velocities measured for the spectra with S/N$\geq$30. The maximum rotation velocity $v_{max}$ is taken as the difference of the most extreme velocity measurements to the system value. The central velocity dispersion $\sigma_{0}$ is taken as the measurement for the spatial aperture closest to the galaxy centre. The anisotropy parameter $(v/\sigma_{0})^{*}$ is defined as $(v_{max}/\sigma_{0})/ \sqrt{\epsilon/1-\epsilon}$. In the formula we used the mean ellipticity value derived from our imaging analysis (see Table~\ref{photfit}).}
\label{kinematics}
\end{table}

\subsection{NGC~1400 results}
\label{1400kin}
The very central regions of NGC~1400 has been studied by many authors (\citealt{tonry81}; \citealt{davies87}; \citealt{faber89}; \citealt{beuing02}). In this study we spatially resolved 82 regions along the major axis of NGC~1400, reaching a maximum radius of 1.3$r_{e}$ from the galaxy centre. The radial profiles are presented in Figure~\ref{kin2} and the kinematic measurements are reported in Table~\ref{kinematics} (all the data are available electronically).

The galaxy displays a net rotation about the minor axis. The velocity rises sharply by $\sim$~100~km~s$^{-1}$ inside $\sim$~1~arcsec (0.10~kpc) from the galaxy's centre and it remains constant out to $\sim$~7~arcsec (0.71 kpc). The $h_{3}$ profile has a typical point-symmetric profile with respect to the centre of the galaxy, characterised by the sign of $h_{3}$ being opposite to the rotation velocity sign. In the $v_{max}/\sigma_{0}$ vs. $\epsilon$ diagram (\citealt{binney82}) the galaxy lies on the line describing models of oblate spheroids with isotropic residual velocities and rotational flattening.
The anisotropy parameter is 1.18~$\pm$~0.12.

\section{SPATIALLY RESOLVED SURFACE PHOTOMETRY}
\subsection{Surface photometry modelling}
The photometric parameters of NGC~1407 and NGC~1400 were extracted using
the ISOPHOTE package in STSDAS. The images were background-subtracted and masked to exclude the gap between the individual chips, as well as any foreground stars and background galaxies. The background subtraction was based on the average of median values from several 10 $\times$ 10 pixel boxes, sampled in regions not affected by the galaxy light. 

The ISOPHOTE package is based on the method of \cite{jed87} and it
involves the creation of a smooth elliptical galaxy model to fit to
the galaxy image (e.g. \citealt{ft92}). The method makes the
assumption that each isophote can be modelled by an ellipse whose
centre, ellipticity and position angle are allowed to vary to reach
the best-fit value; only the length of each semi-major axis (hereafter
radius) is pre-defined and fixed.

We used the technique and the synthetic coefficients from \cite{siri05} to convert from the HST/ACS instrumental F435W and F814W system to the standard Johnson-Cousins system. Final magnitudes were corrected for Galactic extinction using the reddening values
from the DIRBE dust maps of \cite{schl98}. The extinction values for the different filters are: 
$A_{F435W}~=~0.297$ mag, $A_{F814W}~=~0.134$ mag, $A_{g'}~=~0.279$ mag, $A_{r'}~=~0.173$ mag, $A_{i'}~=~0.125$ mag.

The errors in intensity and local gradient are obtained directly from the rms scatter of intensity data along the fitted ellipse. Total errors associated to intensity data points are computed summing in quadrature the above mentioned errors in intensity and the errors in the background subtraction. Errors in the Galactic extinction correction values and in the zero-points are negligible with respect to the errors in the intensity profiles.

\begin{figure*}
\includegraphics[scale=0.5, angle=0]{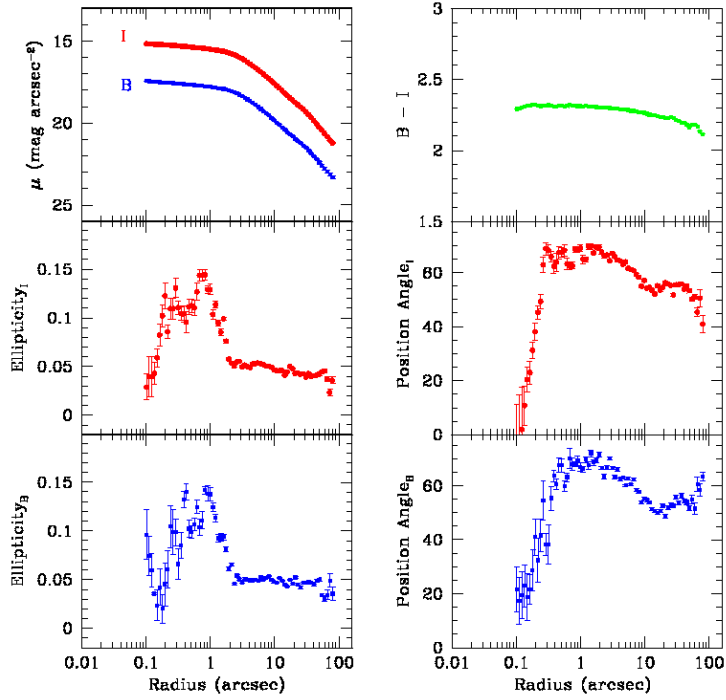}
\caption[]{Spatially resolved radial photometric parameters of NGC~1407.
B (blue filled squares) and I (red filled circles) band surface brightness profiles. Observed B$-$I colour profile (filled green circles). The B and I band isophotes ellipticity and position angle profiles are also shown.}
\label{n1407phot1}
\end{figure*}

\begin{figure*}
\includegraphics[scale=0.5, angle=0]{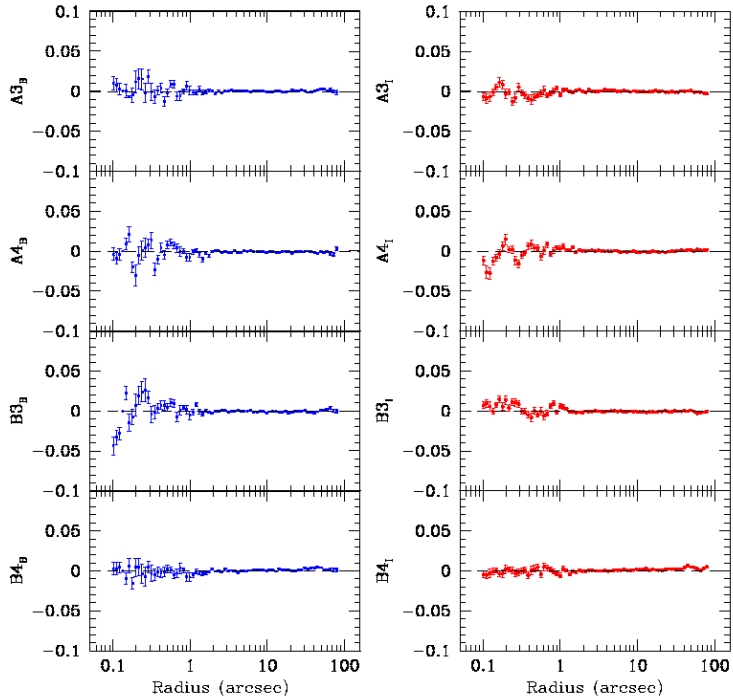}
\caption[Figure2]{NGC~1407 photometric parameters. B (blue filled squares) and I band (red filled circles) 3rd and 4th cosine ($A_{3}$, $A_{4}$) and sine ($B_{3}$, $B_{4}$) terms. Positive values for $A_{4}$ indicate disky isophotes, whereas negative values represent boxy isophotes.}
\label{n1407phot2}
\end{figure*}

\begin{figure*} 
\includegraphics[scale=0.5, angle=0]{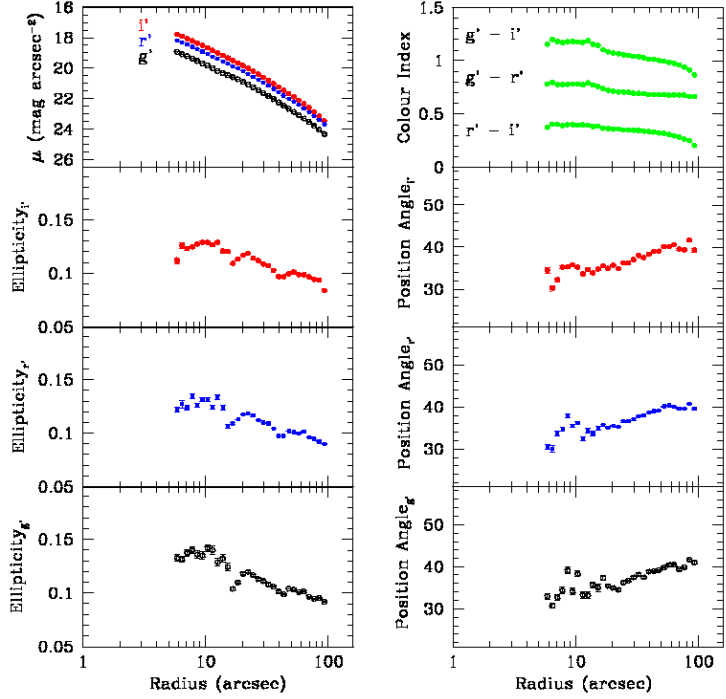}
\caption[]{Spatially resolved radial photometric parameters of
NGC~1400.  g' (black open circles), r' (blue filled squares) and i' (red filled circles)
band surface brightness profiles. Observed g'$-$i', g'$-$r' and
r'$-$i' colour profiles (filled green circles). The g', r' and i' band ellipticity
and position angle profiles are also shown.}  \label{n1400phot1}
\end{figure*}

\begin{figure*}
\includegraphics[scale=0.5, angle=0]{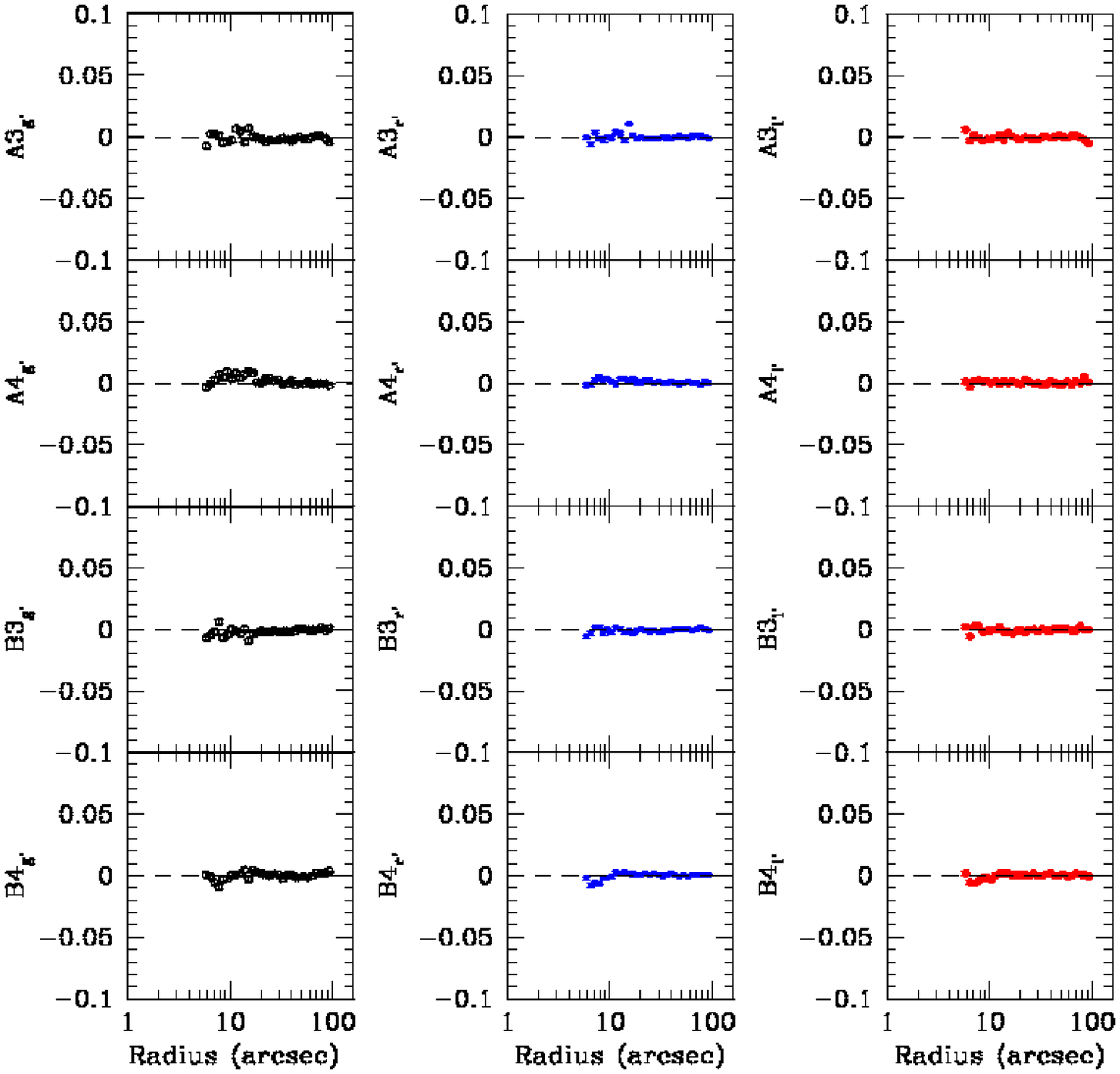}
\caption[Figure2]{NGC~1400 photometric parameters. g' (open black circles), r' (blue filled squares) and i' band (filled red circles) 3rd and 4th cosine ($A_{3}$, $A_{4}$) and sine ($B_{3}$, $B_{4}$) terms. Positive values for $A_{4}$ indicate disky isophotes, whereas negative values represent boxy isophotes.}
\label{n1400phot2}
\end{figure*}

Photometric modelling results are shown in Figures~\ref{n1407phot1}~(NGC~1407) and \ref{n1400phot1}~(NGC~1400). 
The photometric parameters presenting the 3rd and 4th harmonic deviations from ellipses are shown in Figures~\ref{n1407phot2}~(NGC~1407) and \ref{n1400phot2}~(NGC~1400). 

\subsection{Surface brightness profile fitting}
\label{surf}
The surface brightness radial profiles of NGC~1407 were fit to three different galaxy light profiles: de Vaucouleurs $r^{1/4}$ law (\citealt{vac59}) , ``Nuker'' law (e.g. \citealt{lauer95}) and ``core-S\'{e}rsic" law (\citealt{graham03}; \citealt{truj04}). The NGC~1400 radial profiles were only fit to the de Vaucouleurs $r^{1/4}$ law due to the loss of the central regions (see Section~\ref{subaru}). The profile fitting results are reported in Table~\ref{lightfit}.

The Nuker law:

\begin{equation}
  I(r)= I_{b}2^{(\beta - \gamma)/\alpha}(r_{b}/r)^{\gamma}[1+(r/r_{b})^{\alpha}]^{(\gamma-\beta)/\alpha}
\end{equation}

\noindent is an \textit{ad hoc} empirical double power law
used to parametrise the inner 10 arcsec regions of high resolution HST profiles (e.g. \citealt{lauer95}). The surface brightness profiles of early-type galaxies are often
characterised by a double behaviour as a function of the galaxy
radius. They consist of a shallow power law regime at small radii and
a steeper power law at large radii. The Nuker law combines two
power laws with different slopes $\gamma$ and $\beta$ to describe this behaviour. 
The break radius $r_{b}$ indicates the transition between the two regimes and
the parameter $I_{b}$ expresses the surface brightness at $r_{b}$ (e.g., $\mu_{b}$). To evaluate the
variations in the sharpness of the break an additional $\alpha$
parameter has been used. 

\begin{table*}
\begin{center}
\begin{tabular}{cccccccccccccc}
\hline
\hline
Galaxy & Band & Range & Profile model& $\alpha$ & $\beta$ & $\gamma$ & $b$ & $n$ & $r_{b}$ & $r_{e}$ & $\mu_{b}$ & $\overline{\mu_{e}}$& SMBH \\
    (1)& (2)  &  (3)  &   (4)   &  (5)     &   (6)   & (7)      & (8)& (9) & (10) &  (11)  & (12) & (13) & (14)\\
\hline 
NGC~1407 & B  & 2$-$100   & de Vaucouleurs  &  --  &  --  &   -- & --   & --    & --   & 71.23 &  --   & 20.87 & --\\ 
         &    & 0.1$-$10  & Nuker           & 2.71 & 1.37 & 0.12 & --   & --    & 2.68 & --    & 18.22 & -- & --\\
         &    & 0.1$-$100 & Core-S\'{e}rsic & 3.01 & --   & 0.12 & 14.26& 8.28  & 2.33 & 67.35 & 18.08 & & 1.02\\ 

         &   I   & 2$-$100   & de Vaucouleurs  &  --  &  --  &   -- & --   & --    & --   & 72.55 &  --   & 19.65 &  \\ 
         &    & 0.1$-$10  & Nuker           & 2.68 & 1.41 & 0.12 & --    & --   & 2.68 & -- & 15.94    & --    &    \\
         &    & 0.1$-$100 & Core-S\'{e}rsic & 2.94 & --   & 0.12 & 15.09 & 8.42 & 2.37 & 72.59 & 15.82 &       & 1.05\\
\hline
NGC~1400& g'  & 5$-$100  & de Vaucouleurs    & --   & --   & --   & --   & --    & --   & 26.58    & -- & 21.44 &--  \\
        & r'  & 5$-$100  & de Vaucouleurs    & --   & --   & --   & --   & --    & --   & 27.21    & -- & 20.74 & -- \\
        & i'  & 5$-$100  & de Vaucouleurs    & --   & --   & --   & --   & --    & --   & 27.80    & -- & 20.39 & -- \\
        & V$^{\dag}$  & 0.02$-$10  & Nuker   & 1.39 & 1.32 & 0.00 & --   & --    & 0.33 & --       & -- & --& --\\
\hline
\hline
\end{tabular}
\end{center}
\caption{Surface brightness fitting results. Col~1: name of the galaxy. Col~2: photometric band. Col. 3: radial range in arcsec. Col.~4: galaxy light profile model. Col.~5--12: parameters of the profile models, see text for details; $r_{b}$ and $r_{e}$ in arcsec; $\mu_{b}$ in mag arcsec$^{-2}$. Col.~13: mean surface brightness within $r_{e}$ in mag arcsec$^{-2}$; from $\overline{\mu}_{e}$~=~2.5log~($\pi r_{e}^{2}$)~+~$m(r_{e})$ (\citealt{bingelli84}; \citealt{graham97}), $m(r_{e})$ is the enclosed magnitude within $r_{e}$ estimated from the surface photometry modelling. Col.~14: estimate of the supermassive black hole mass, expressed in $\times$~10$^{9}$M$_{\sun}$, from the $M_{bh}-n$ relation of \citet{graham07}; the error estimates are reported in Section~\ref{surf}. The symbol $\dag$ indicates results from \citet{faber97}.}
\label{lightfit}
\end{table*}

The core-S\'{e}rsic law is a profile model proposed to better understand the core/power-law dichotomy (\citealt{faber97}) found in the radial surface-brightness profiles of early-type galaxies.
Briefly, this is done by modifying the Nuker law through the inclusion of a S\'{e}rsic model to describe the outer part of the profile. The new law has the ability to model the entire galaxy radial profile. The model is:
\begin{equation}
I(r) = I'[1+ (r_{b}/r)^{\alpha}]^{\gamma/\alpha}exp{{-b[(r^{\alpha}+r^{\alpha}_{b})/r_{e}^{\alpha}]^{1/(n\alpha)}}},
\end{equation}
where 
\begin{equation}
I'=I_{b}2^{-\gamma/\alpha}exp[b2^{1/\alpha n}(r_{b}/r_{e})^{1/n}],
\end{equation}
and the other parameters have the same general meaning as in the Nuker or S\'{e}rsic laws: $n$ is the shape parameter of the outer S\'{e}rsic part, $r_{e}$ is the effective half-light radius of the outer $r^{1/n}$ profile. The quantity $b$ is defined to be a function of the parameters $\alpha, r_{b}/r_{e}$ and $\gamma$ such that $r_{e}$ becomes the radius enclosing half of the light of the galaxy model. 

The flat core profile observed in the surface brightness of NGC~1407 may be caused by the coalescence of merging supermassive black holes (SMBHs) that partially evacuated the inner region of the galaxy (e.g \citealt{faber97}; \citealt{milo01}); NGC~1400 is already known to be a core galaxy (\citealt{faber97}; see also Table~\ref{lightfit}). Conversely, a steep power-law profile has been proposed to be the consequence of adiabatic growth of the central black hole, reshaping the inner galaxy region (e.g. \citealt{van99}). Recently, \cite{graham07} investigated the possibility of a high order $M_{bh}-n$ relation finding an empirical log-quadratic relation for predicting SMBH masses. The relation is
\begin{equation}
log(M_{bh}) = 7.98 + 3.70log(\frac{n}{3}) - 3.10[log(\frac{n}{3})]^{2}
\end{equation}
where $n$ is the S\'{e}rsic index, expressing a measure of the stars concentration within the bulge of the galaxy. We applied this formula adopting the $n$ values from the core-S\'{e}rsic law best-fit to the $\mu_{B}$ and $\mu_{I}$ surface brightness radial profiles of NGC~1407. The maximum error on the estimated black hole mass is obtained using equation (8) from \cite{graham07}. We find a SMBH mass of 1.02~$\pm$~0.03~$\times$~10$^{9}$~$M_{\sun}$ and 1.05~$\pm$~0.01~$\times$~10$^{9}$~$M_{\sun}$, respectively. The mass of a central SMBH is known to be connected with properties of the host galaxy. Our SMBH mass estimation is consistent within 1$\sigma$ of the $M_{bh}-\sigma$ relation (\citealt{ferrarese00}).

\subsection{Isophotal shape parameters and fine structure}
The position angle of NGC~1407 and NGC~1400 shows radial variation, twisting respectively of $\sim$~20$^{\circ}$ and $\sim$~10$^{\circ}$ with the galactocentric radius. The ellipticity profiles of NGC~1407 do not present significant radial variation. Some scatter is found for radii smaller than 1 arcsec. In contrast, the ellipticity of NGC~1400 varies from $\sim$~0.14 in the centre to $\sim$~0.09 at larger radii. Mean values are presented in Table~\ref{photfit}, together with results from previous works.

We adopt the same definition of boxiness proposed by \cite{faber97}: i.e. disky isophotes are
defined by positive values of A$_{4}$ while boxy isophotes are indicated by negative A$_{4}$ values. The A$_{4}$ radial profiles of both galaxies appear uniformly equal to zero. Small scatter is detected in the very inner regions of the NGC~1407 profile. Consequently, the isophotal shape of the two galaxies is classified as neutral.

Using the BMODEL task in IRAF we created smooth two-dimensional models of the elliptical structure of the galaxies. Residual images were obtained, by subtracting these models from the original images, to create a better contrast between the fine structure and the smooth background brightness profile of the underlying galaxy. The residual images appear smooth and regular and we do not find any features in the central or external regions.

\section{The peculiar velocity of NGC~1400}
We examine the possibility of an interaction between NGC~1407 and NGC~1400. In Section~2 we have seen that the two galaxies lie at almost the same distance (see Table~\ref{distance}), with a remarkable difference in velocity, $\Delta v$~$\sim$~1100~km~s$^{-1}$. A reasonable hypothesis is that NGC~1400 is falling into the group, and may be experiencing some sort of interaction with the brightest group galaxy NGC~1407 or the intergalactic medium. Recently, \cite{trentham06} identified a second fainter (M$_{B}$~=~$-$14.78~mag) galaxy (dwarf elliptical LEDA~074880) with a peculiar velocity of $\sim$~940~km~s$^{-1}$; this may suggest an infalling subgroup associated with NGC~1400. 

\cite{osullivan01} and \cite{osmond04} analysed the X-ray emission properties of the two galaxies and of the group IGM, respectively, from ROSAT PSPC pointed observations. The former found that NGC~1407 and NGC~1400 lie well inside 1$\sigma$ scatter on the $L_{X} - L_{B}$ relation for early-type galaxies. The latter, using a $\beta$-model found a total X-ray luminosity of $\sim$~10$^{42}$~erg~s$^{-1}$ within 570 kpc, placing the NGC~1407 group directly on the $L_{X}-T_{X}$ and $L_{X}-\sigma_{v}$ relations. \cite{tully06} claimed to observe a Bondi-Hoyle (\citealt{bondi44}) wake in the X-rays around NGC~1400 and hence evidence for a physical interaction, but there are many low contour extensions so it is not clear if this interpretation is correct.

In terms of scaling relations we find that both galaxies follow the Faber-Jackson (\citealt{faber76}), fundamental plane (\citealt{djo87}; \citealt{dressler87}), $M_{bh}-\sigma$ (\citealt{ferrarese00}) and $L_{X} - L_{B}$ relations, within the normal scatter. \cite{forbes98} showed that the scatter about the fundamental plane is correlated with the galaxy age and the position of the galaxy relative to the plane is dependent on the time since its last starburst. The measured residuals for NGC~1407 and NGC~1400 are positive, corresponding to galaxies with an age $\geq$~10~Gyr. This is consistent with the results of Paper~II; they found that both galaxies are uniformly old with no indication of young stars at the galaxy centre. The radial kinematics show no obvious evidence for disturbance in the outer parts. There is no sign of tidal distortion in the direct images or from our galaxy isophote modelling. 

\begin{table}
\begin{center}
\begin{tabular}{cllccccc}
\hline
\hline
Galaxy     & Band & Range &  $\overline{\epsilon}$ &  $\overline{P.A.}$ \\
  (1)      &  (2) &  (3)  &  (4)                   &     (5)            \\
\hline
NGC 1407   &  B  & 0.1 $-$ 100            &   0.05 $\pm$ 0.01 &  58.4 $\pm$ 1.2 \\
           &  I  & 0.1 $-$ 100            &   0.05 $\pm$ 0.01 &  58.3 $\pm$ 0.9 \\
           &  B$^{\dag}$  & 1 $-$ 110     &   0.05 $\pm$ 0.01 &  60.3 $\pm$ 1.2 \\ 
           &  I$^{\dag}$  & 1 $-$ 110     &   0.05 $\pm$ 0.01 &  59.6 $\pm$ 1.0 \\ 
           &  V$^{\dag}$  & 1 $-$ 110     &   0.05 $\pm$ 0.01 &  59.8 $\pm$ 1.1 \\ 
\hline
NGC 1400   & g'    & 5 $-$ 100     & 0.11 $\pm$ 0.01       &  37.0 $\pm$ 0.6  \\
           & r'    & 5 $-$ 100     & 0.11 $\pm$ 0.01       &  36.1 $\pm$ 0.4  \\
           & i'    & 5 $-$ 100     & 0.11 $\pm$ 0.01       &  36.1 $\pm$ 0.3  \\
           & V$^{\dag \dag}$ & 0.02 $-$ 10 &  0.19 $\pm$ 0.01 &  30.2 $\pm$ 0.3  \\
\hline 
\hline
\end{tabular}
\end{center}
\caption{Galaxy isophote modelling results. Col~1: name of the galaxy. Col~2: photometric band. Col. 3: radial range in arcsec. Col.~4: mean ellipticity. Col.~5: mean position angle in degree. The symbol $\dag$ indicates results from \citet{faber97}. The symbol  $\dag \dag$ indicates results from \citet{lauer95}.}
\label{photfit}
\end{table}

\section{Summary}
This is the first paper of a series with the aim of studying the star formation and evolutionary history of the two early-type galaxies NGC~1407 and NGC~1400. Here, we have presented spatially resolved radial kinematics and surface photometry. The spectroscopic analysis is performed using high signal-to-noise long-slit data obtained at the ESO 3.6m telescope. The imaging study is based on high-resolution multi-band data from the HST/ACS and wide-field imaging from Subaru Suprime-Cam. The high-quality of the data allowed us to focus our study on the properties of spatially resolved galactic regions and to trace the overall radial trends. The radial profiles of the kinematic parameters $v$, $\sigma$, $h_{3}$ and $h_{4}$ have been recovered. From the imaging analysis we obtained surface brightness and colour index profiles. We also measured the radial profiles of the isophotal shape parameters and created residual galaxy images to search for fine structure.

\begin{list}{}{}
\item \textbf{NGC~1407.} The kinematic study suggests a rotationally supported galaxy (or marginally anisotropic) with the presence of a possible kinematically decoupled core; the detection is uncertain and potentially may be caused by a misalignment of the slit with the nucleus. The surface brightness profiles reveal a flat core probably caused by the presence of a central supermassive black hole, with an estimated mass of $\sim$1.03~$\times$~10$^{9}$~$M_{\sun}$. We find NGC~1407 to be an elliptical galaxy (A$_{4}$=0) with a mean position angle of 59$^{\circ}$ and a small ellipticity value of 0.05. No fine structure is detected in the residual images.

\item \textbf{NGC~1400.} The anisotropy parameter and the kinematic profiles indicate NGC~1400 to be rotationally supported with evident minor axis rotation and flattening due to fast rotation. The galaxy was already found to have a flat core in the surface brightness profile (\citealt{lauer95}). The galaxy isophote modelling shows elliptical isophotes at all radii (A$_{4}$=0) and no fine structure is detected in the residual galaxy images. The ellipticity is more pronounced, 0.11, and the position angle is 36$^{\circ}$. From our results we would tend to classify NGC~1400 as an E1 galaxy, in contrast with the E$-$S0 morphological classification proposed in literature. 
\end{list}

We found no evidence to support an interaction between NGC~1400 and NGC~1407 in our data. We speculate that what we are now witnessing might be the first infall of a subgroup, dominated by the NGC~1400 galaxy, into the NGC~1407 group. 

\section{ACKNOWLEDGEMENTS}
Based on observations made with the NASA/ESA Hubble Space Telescope, obtained from the Data Archive at the Space Telescope Science Institute, which is operated by the Association of Universities for Research in Astronomy, Inc., under NASA contract NAS 5-26555. These observations are associated with program ID=9427 PI=Harris. We thank Lee Spitler for providing the Suprime-Cam photometric data. We also thank Aaron Romanowsky for useful discussions on surface brightness profile fitting. DF, RP and SB thank the ARC for financial support.
 
\end{document}